\documentclass{iopconfser}

\usepackage{hyperref}
\hypersetup{
	colorlinks=true,
}

\usepackage{graphicx}
\usepackage{subcaption}
\graphicspath{{.}{./figures/}{../../figures/}}

\usepackage[print-unity-mantissa=false]{siunitx}
\DeclareSIUnit{\Sa}{Sa}
\DeclareSIUnit{\GiB}{GiB}
\DeclareSIUnit{\MiB}{MiB}
\DeclareSIUnit{\Mbps}{Mbps}

\usepackage{tikz}
\usetikzlibrary{positioning,arrows.meta}

\usepackage[backend=biber,style=numeric,sorting=none,sortcites=true,maxbibnames=1]{biblatex}
\DeclareSourcemap{
 \maps[datatype=bibtex]{
  \map{
    \step[fieldsource=collaboration, final=true]
    \step[fieldset=usera, origfieldval, final=true]
  }
 }
}

\newbibmacro*{collaboration}{%
	\iffieldundef{usera}
	{}
	{%
		\setunit{\addspace}%
		\addspace
		for the \printfield{usera} Collaboration%
	}%
}

\renewbibmacro*{author}{%
	\printnames{author}%
	\usebibmacro{collaboration}%
}
\begin{document}

\title{BOLT: Imaging Lightning and Terrestrial Gamma-ray Flashes at the Pierre Auger Observatory}

\author{Eric-Teunis de Boone$^{1}$ for the Pierre Auger Collaboration$^{2}$}

\affil{$^1$Center for Particle Physics Siegen, Department Physik, Universit\"at Siegen, Germany\\

$^2$Observatorio Pierre Auger, Av. San Mart\'in Norte 304, 5163 Malargu\"e, Argentina}

\email{
	\href{mailto:spokespersons@auger.org}{spokespersons@auger.org}\\
	Full author list: \url{https://www.auger.org/archive/authors_2026_07.html}
}

\begin{abstract}
	Lightning-related phenomena are known to interact with and influence all detector systems of the Pierre Auger Observatory.
	Notably, the Surface Detector has recorded unique signals associated with Terrestrial Gamma-ray Flashes (TGFs),
		which are rare phenomena linked to the initial processes of lightning.
	Studying TGFs in coincidence with their parent lightning discharges helps to constrain the mechanisms responsible for their production,
		providing key insights into how these high-energy emissions are generated and evolve within thunderstorms.

	To image the parent lightning discharges, we are developing the Broadband Observatory of Lightning and TGFs (BOLT),
		an interferometric lightning mapping array that enhances the unique capabilities of the Observatory by providing high-resolution 3D (spatial and temporal) imaging of lightning propagation.
	BOLT consists of VHF radio detectors that have been previously developed for the Auger Engineering Radio Array, sensitive in the $\SIrange{30}{80}{\MHz}$ range.
	The array is co-located with the Surface Detector to enable direct correlation between lightning activity and candidate TGF signals at the Pierre Auger Observatory.

	The BOLT array is in commissioning, with successful station firmware modifications to acquire and stream multi-second waveforms.
\end{abstract}

\section{Introduction}
\label{sec:intro}
	The Pierre Auger Observatory is a hybrid cosmic-ray observatory based near Malargu\"e, Argentina.
	It is designed to detect and study ultra-high-energy cosmic rays with energies above \SI{1e17}{\eV},
		covering \SI{3000}{\km\squared} with its Surface Detector.
	The sensitive detector systems of the Observatory are affected by thunderstorms and other lightning-related phenomena.
	For example, the Surface Detector has recorded unique signals associated with Terrestrial Gamma-ray Flashes (TGFs)~\cite{PierreAuger:2025yti}.

	TGFs are short bursts of gamma-ray photons linked to the initial stages of lightning discharges.
	The first observations of TGFs were made by the BATSE instrument on the Compton Gamma-ray Observatory satellite~\cite{Fishman:1994}.
	Production mechanisms are not yet fully understood.
	Studying TGFs in coincidence with their parent lightning discharges
		can therefore provide constraints on the processes responsible for their production.

	Stepped leaders and other lightning development stages emit impulsive VHF radiation.
	VHF observations therefore provide a means of reconstructing the development of lightning channels.
	Although the Auger~Engineering Radio~Array (AERA) was not designed for it,
		such signals have also been observed in AERA data~\cite{Weitz:2024jou}.

	The Broadband Observatory of Lightning and TGFs (BOLT) therefore aims to reuse AERA hardware to reconstruct the lightning development.

	\paragraph{}
	This work describes the implementation of BOLT,
		focusing on the modifications required to extend the AERA station readout from short air-shower traces to multi-second lightning observations,
		and reports on the commissioning status.
	In addition,
		self-triggered AERA data has been analysed for coincidence with data from the Lightning Detection System
		to identify further lightning candidate events,
		as proposed in~\cite{Weitz:2024jou},
		to investigate the dynamic range for lightning observations above the Pierre~Auger Observatory.

\section{BOLT Array Configuration and Hardware}
\label{sec:hardware}
	The BOLT array repurposes radio detectors previously developed for AERA,
		sensitive in the $\SIrange{30}{80}{\MHz}$ frequency range.
	The array comprises 11 stations distributed over three clusters with increasing baselines: Core, Medium-range and Remote~\cite{Weitz:2024jou}.
	The stations are designed to receive a trigger from the Lightning Detection System~\cite{Rautenberg:2015jvj}.
	In later stages, a trigger from the Surface Detector is foreseen.

\begin{figure}
	\begin{subfigure}{0.7\textwidth}
		\includegraphics[width=\textwidth]{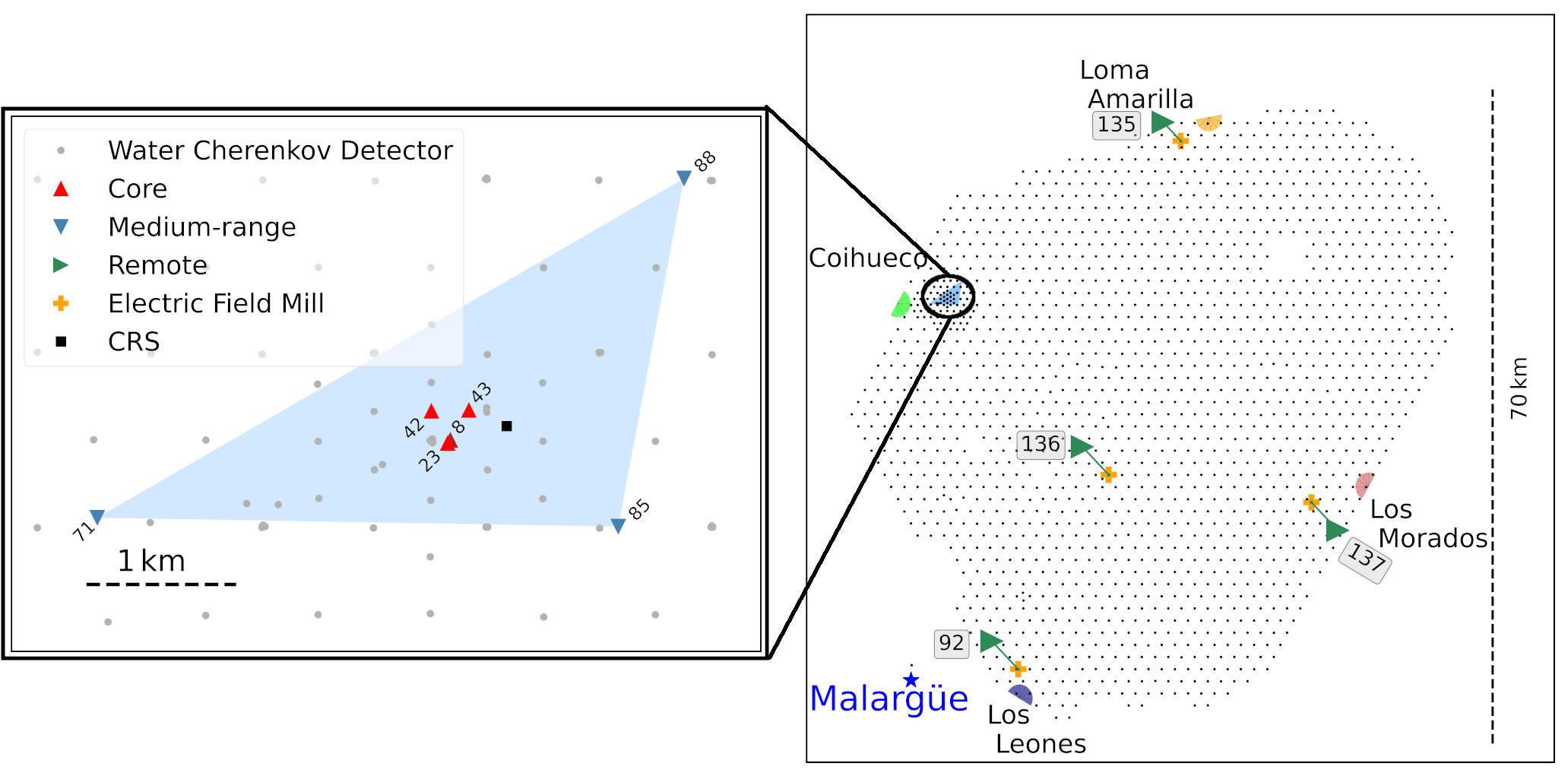}
		\caption{}
		\label{fig:array_configuration}
	\end{subfigure}
	\hfill
	\begin{subfigure}{0.23\textwidth}
		\includegraphics[width=\textwidth]{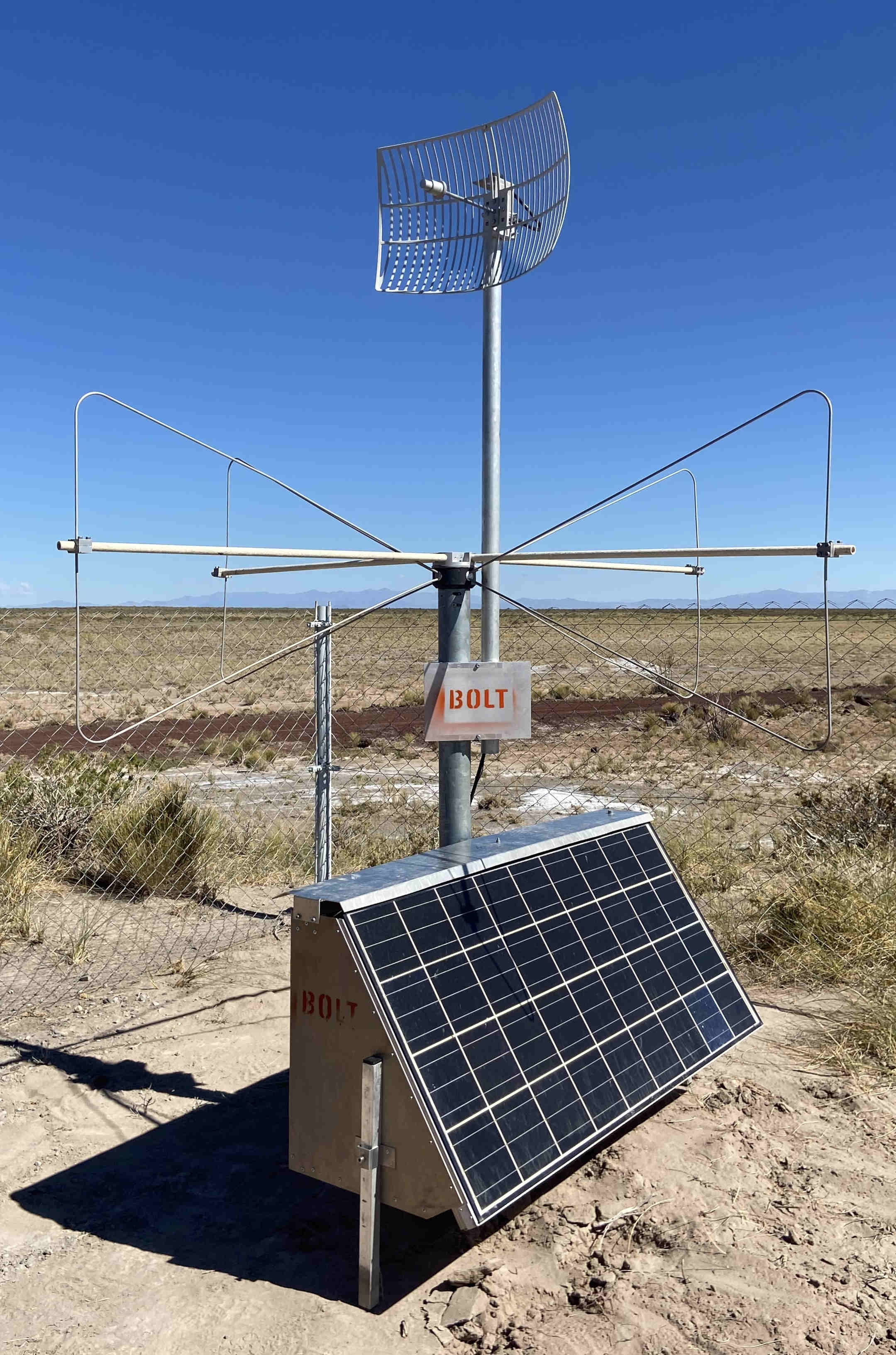}
		\caption{}
		\label{fig:BOLT137}
	\end{subfigure}
	\caption{
		Left: The configuration of the BOLT stations within the Pierre Auger Observatory.
		Color-coded triangles mark the station and its cluster: Core (red), Medium-range (blue) and Remote (green).
		Light-grey circles mark the positions of Surface Detector stations.
		Right: Photograph of BOLT137 (a Remote station) near Los Morados after deployment in Spring 2026.
	}
	\label{fig:bolt_array_and_station}
\end{figure}

	Figure~\ref{fig:array_configuration} presents the different clusters
		in relation to the Surface Detector and the Fluorescence Detector locations.
	The Core and Medium-range clusters have baselines from \SIrange{58}{127}{\m} and \SIrange{1}{2.5}{\km}, respectively.
	The Remote cluster has baselines up to \SI{60}{\km}.

	The positions in the Core and Medium-range clusters coincide with former AERA positions
		and the hardware is unchanged except for the digitizer.
	For the Remote cluster, four complete AERA stations have been relocated.

	The three clusters address different requirements for the lightning reconstruction.
	The short baselines of the Core cluster help to reduce pulse confusion,
		which can occur when the relative arrival times of a set of pulses are reordered between stations due to the propagation geometry.
	The larger baselines of the Medium-range cluster improve the spatial resolution of reconstructed sources,
		while the Remote cluster improves both the spatial resolution and sensitivity.

	\paragraph{}
	Each station employs two orthogonal dipole antennas
		aligned to the geomagnetic north-south and east-west directions.
	All stations employ the Butterfly antenna type (see Figure~\ref{fig:BOLT137}),
		except for BOLT008 which employs a Log-Periodic Dipole Antenna.

	Internally, the BOLT stations use the AERA phase II and phase III digitizers designed by KIT/BUW\footnote{Karlsruhe Institute of Technology / Bergische Universit\"at Wuppertal}.
	The main features of these digitizers are four 12-bit analog-to-digital converters (ADC)
		running at a sampling rate of \SI{180}{\mega \Sa \per \s}
		and a \SI{4}{\GiB} sampling buffer,
		while operating at a design power consumption under \SI{15}{\W}.
	Depending on the number of enabled ADCs, the \SI{4}{\GiB} sampling buffer provides a sampling depth of \SIrange{4}{16}{\s}.

	The KIT/BUW digitizer is operated by an Altera Cyclone III FPGA
		hosting a uClinux-based system for coordinating higher-level communications and DAQ software.
	Communications are provided through long-range directional Wi-Fi or an optical fiber connection.

	Absolute timing is derived from an i-Lotus~M12M GPS module.
	It provides a 1~Pulse-Per-Second (PPS) signal used as the timing reference for the digitizer hardware.
	A serial connection is available for communications between the uClinux and GPS module.

	The digitizer interfaces with an analog filterboard that delivers the power for the low-noise amplifier incorporated in the antennas.
	Two of the four output channels have an additional gain of \SI{20}{\decibel} for increased sensitivity.

\vfill
\section{Adapting AERA DAQ for BOLT}
\label{sec:daq}

	Since AERA is designed to measure the short radio signals emitted by cosmic-ray extensive air showers,
		which are typically a few nanoseconds long,
		the station DAQ pipeline is designed for trace lengths of approximately \SI{11}{\us}.
	The FPGA continuously reads the ADC values into the sample buffer,
		and, on a trigger (either internal or external),
		the relevant samples are read out by the station DAQ software.
	With the finite buffering depth,
		it is therefore possible that a trigger arrives too late and relevant data are already out of the memory and the trigger is missed,
		for example because of communication disruptions or a high number of triggers accumulating readout latencies.

	\paragraph{}
	The main challenge for a BOLT station to record a complete lightning event is therefore the increase of the trace lengths to the order of seconds.
	This increases the amount of waveform data per event to the \SI{}{\GiB} range.

	As the uClinux has \SI{128}{\MiB} of available memory,
		such an event can no longer be stored internally and needs to be streamed directly to external storage.
	With a digitizer readout bandwidth under \SI{15}{\Mbps}, a readout of \SI{1}{\GiB} takes over \SI{9}{\min}. 

	Therefore, the main FPGA modification implemented for BOLT is the ability to stop the continuous sampling.
	This also makes the system more robust against communication disruptions as an event remains available in the sampling buffer until the sampling is restarted.

	Because of this sampling ``freeze'' and the streaming approach, new DAQ software needed to be designed.
	Finally, a storage backend accepting the streaming from the stations is required.

	\subsection{FPGA and station DAQ}
	The Data Management Unit (DMU) module in the FPGA is responsible for
		both storing incoming waveform data in the SODIMM memory
		as well as servicing the timestamp-based read requests from the uClinux driver.
	Multiple modifications inside the DMU now suspend the writing process
		after the existing \verb|REGFILE_START_STOP| signal is toggled,
		preserving the full buffer content for subsequent readouts.

	The original implementation imposed strict validity checks on the data request to prevent invalid data access.
	For the long trace readout, these restrictions prevent access to earlier data in the buffer and must be relaxed.
	Address calculation logic has therefore been rewritten to allow access to arbitrary timestamps within the ringbuffer.

	The \verb|REGFILE_START_STOP| signal is available to the uClinux driver which now supports changing the readout mode on-the-fly;
	a Linux application can control the desired behaviour using an \verb|ioctl| call to the driver.
	When enabled, the driver toggles the signal on the first read request causing the freezing behaviour in the DMU.

	The modified firmware has been tested in both modes of operation,
		with long trace lengths ranging from tens of microseconds to over two seconds,
		and is in operation on the modified BOLT stations.

	\paragraph{}
	To handle the large waveforms, a new station DAQ had to be designed.
	A key difference from the original AERA station DAQ is the separation of event information and waveform data into separate ringbuffers,
		allowing the waveform data to be streamed independently of the event control information.

	The program is written against uClibc version v0.9.29-nios2\footnote{GNU C99, GCC 4.1.2} and uses five different pthreads.
	A single orchestrator thread routes a trigger request together with the relevant timing information to the hardware interaction thread.
	This then requests the data from the driver and fills the ringbuffers while the writer thread consumes and streams to external storage.
	A timing thread collects information from the GPS module and the DMU hardware to translate trigger timestamps into hardware timestamps.
	This is needed as the PPS emitted by the GPS module is gated by its oscillator, affecting the PPS placement.

	In typical BOLT operation, reading out the waveform data for a single large event takes tens of minutes.
	The DAQ therefore supports aborting data acquisition and streaming independently, to prevent unnecessary deadtime and bandwidth usage on incidents.
	Depending on the requested trace length, the software can automatically select between the conventional readout mode and the sampling-freeze mode.

	Before the sampling is started, the relevant ADC channels can be configured, allowing the dynamic range and sensitivity to be adapted to the intended measurement.
	For BOLT, channel saturation for nearby lightning events remains under investigation.

	The station DAQ is operational on BOLT stations, but remains under active development.

	\subsection{Data handling and storage}
	To reduce the impact of bandwidth limitations in the network while streaming the large data volumes from a station,
		the external storage must be close to each station.
	The Core and Medium-range clusters share the same external storage, hosted in the Central Radio Station (CRS).
	This is one of the central points of operations for AERA and the BOLT stations connect directly to its network.
	For the Remote cluster, each station has its own external storage computer.
	At a later stage, the data is then transferred to a central storage.

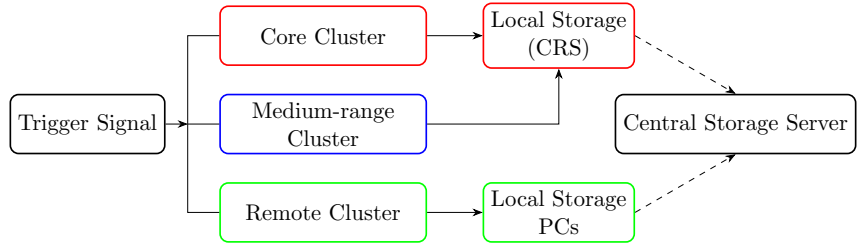
\begin{figure}
	\centering
	\resizebox{0.7\textwidth}{!}{%
	\begin{tikzpicture}[node distance = 4cm, on grid, auto]
		\tikzset{
			node/.style = {
				minimum size = 1cm, align=center, rounded corners, thick, draw=black
			},
			core/.style   = {draw=red},
			medium/.style = {draw=blue},
			remote/.style = {draw=green},
		}

		\node (moni)   [node] {Trigger Signal};

		\node (medium) [node, right =	   of moni,   minimum width=3.5cm, medium] {Medium-range\\ Cluster};
		\node (core)   [node, above = 1.5cm of medium, minimum width=3.5cm, core]   {Core Cluster};
		\node (remote) [node, below = 1.5cm of medium, minimum width=3.5cm, remote] {Remote Cluster};

		\node (core-storage)   [node, right = of core,   core]   {Local Storage\\(CRS)};
		\node (remote-storage) [node, right = of remote, remote] {Local Storage\\PCs};

		\node (central-storage) [node, right = 7cm of medium] {Central Storage Server};

		\coordinate (trigout) at (1.7, 0);
		\draw[arrows=-Stealth] (moni.east) -- (trigout);
		\draw (trigout) -- (medium.west);
		\draw (trigout) -- (1.7, 1.5)  -- (core.west);
		\draw (trigout) -- (1.7,-1.5) -- (remote.west);

		\draw[arrows=-Stealth] (core.east) -- (core-storage.west);
		\draw[arrows=-Stealth] (medium.east) -- (8,0) -- (core-storage.south);

		\draw[arrows=-Stealth] (remote.east) -- (remote-storage.west);

		\draw[arrows=-Stealth, dashed] (core-storage.east)   -- (central-storage.north);
		\draw[arrows=-Stealth, dashed] (remote-storage.east) -- (central-storage.south);
	\end{tikzpicture}
	}
	\caption{
		Dataflow from trigger to long-term storage.
		The Core and Medium-range cluster share the same external ``Local Storage''.
		In the Remote cluster, each station has an individual external ``Local Storage''.
		The data is then moved at low-priority from the ``Local Storage'' to a central storage.
	}
	\label{fig:dataflow}
\end{figure}

	Figure~\ref{fig:dataflow} summarises the dataflow, from trigger to long-term storage.

\vfill
\section{Commissioning}
	The BOLT Core and Medium-range clusters were deployed from autumn 2024 until autumn 2025 and are now in the late stages of commissioning.

	The Remote cluster was deployed in spring 2026 by relocating 4 AERA stations and is in the early stages of commissioning, including establishing the network and storage infrastructure.

\begin{figure}
		\centering
		\includegraphics[height=9cm]{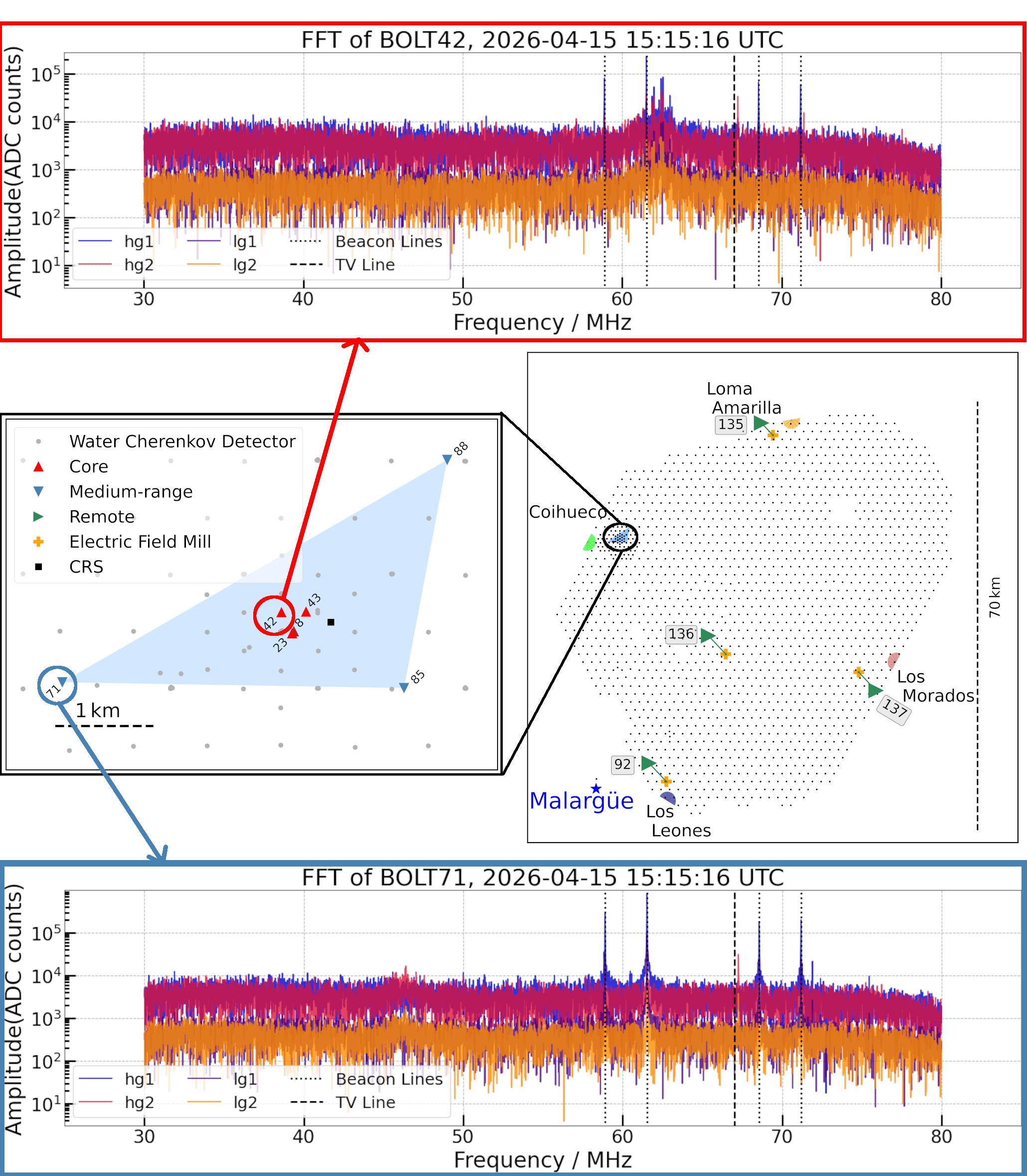}
		\caption{
			Background frequency spectra of BOLT42 and BOLT71 on 2026-04-15 15:15:16 UTC with both low- and high-gain channels.
			The dotted lines indicate the four frequencies of the AERA beacon located at Coihueco used for improving the relative time synchronisation between AERA stations.
		}
		\label{fig:commissioning_202604}
	\end{figure}

	\paragraph{}
	Regular coincident readouts of background data are being performed with the Core and Medium-range stations.
	Figure~\ref{fig:commissioning_202604} shows the frequency spectra of two such coincident traces from BOLT42 (Core) and BOLT71 (Medium-range).
	The indicated beacon lines are known frequencies emitted by a beacon transmitter at Coihueco to improve the relative time synchronisation between stations in AERA to below \SI{2}{\ns}~\cite{PierreAuger:2015aqe}.
	Demonstrating similar synchronisation accuracy for BOLT is part of the ongoing commissioning.

	The figure also shows the effect of the additional \SI{20}{\decibel} gain for the high-gain channels.

	Traces of up to \SI{1}{\s} have been successfully acquired and streamed to the CRS from selected stations,
		demonstrating the complete long-trace readout chain under field conditions.

	In parallel, commissioning includes validation of autonomous station operation and maintenance under field conditions.

\section{AERA Observations of a Lightning Candidate}
	To investigate both the dynamic range and the initial reconstruction approach for lightning signals in AERA,
		existing AERA measurements were searched for events coincident within one second with events reconstructed by the Lightning Detection System~(LDS) at Auger~\cite{Weitz:2024jou}.
	Due to a trigger strategy change that suppresses repeated source locations,
		only few lightning candidate events have been found.

\begin{figure}
	\centering
	\includegraphics[height=8cm]{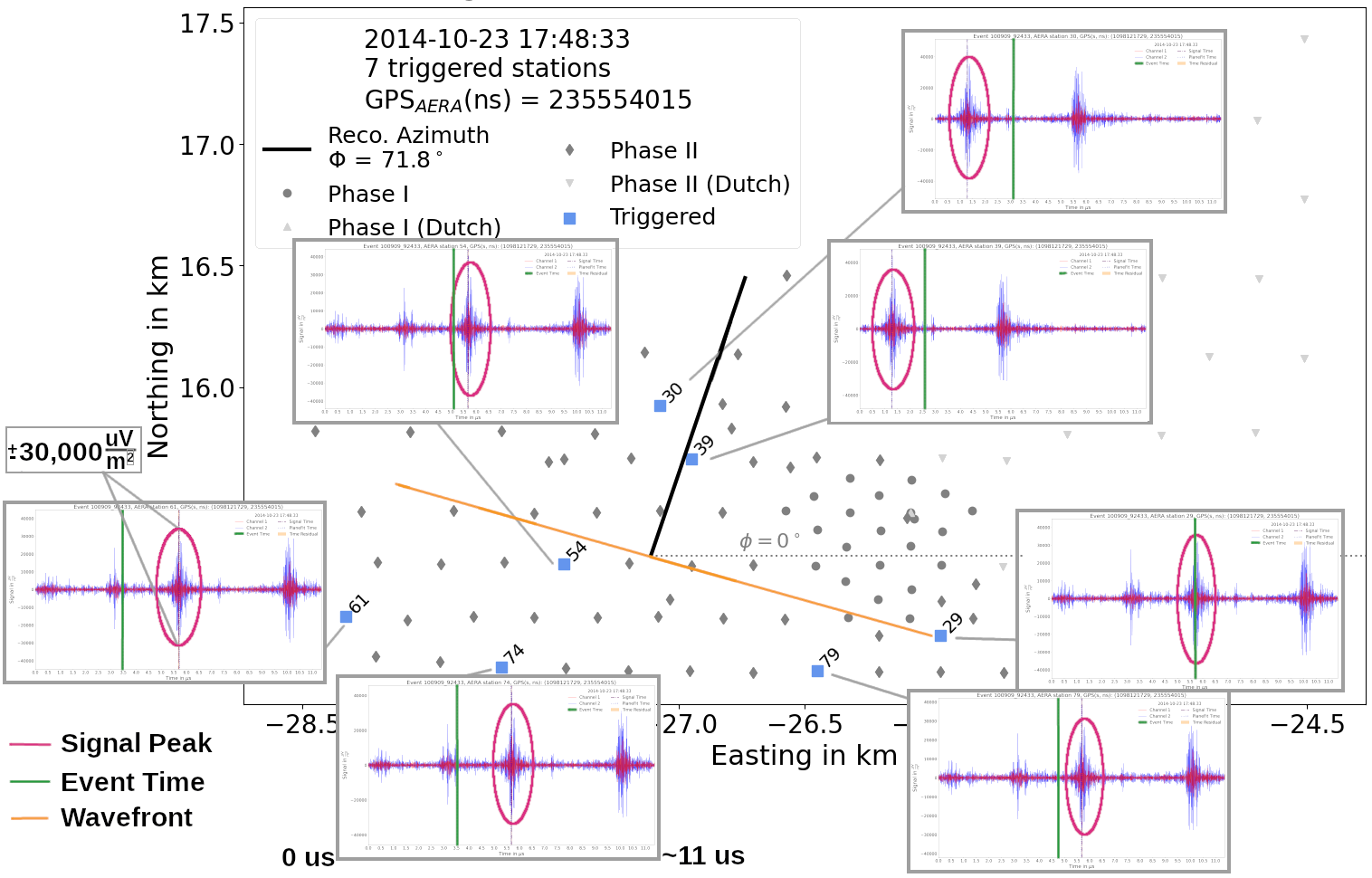}
	\caption{
		AERA lightning candidate event:
		self-triggered AERA measurements within \SI{641}{\ms} of an LDS event, underlaid with the station positions.
		Inside the waveform plots, green vertical lines indicate the same absolute GPS timestamp.
		The pulse circled in red is visible in all stations.
		The black line indicates the reconstructed direction obtained from a plane wave time of arrival reconstruction.
	}
	\label{fig:lightning_candidate_event}
\end{figure}

	Figure~\ref{fig:lightning_candidate_event} shows the clearest candidate,
		with self-triggered AERA measurements recorded within \SI{641}{\ms} of an LDS event.
	Three similar impulsive signal structures are visible in the waveforms of the participating stations.
	Cross-correlation beamforming of the earliest pulse yields a direction consistent with the direction reconstructed by the LDS.
	This agreement suggests that the AERA signals may originate from the same thunderstorm as the event observed by the LDS.

	However, the large distance to the reconstructed LDS position prevents this event from providing a meaningful constraint on the dynamic range for nearby lightning observations.

\section{Status and Outlook}
	The Pierre Auger Observatory has an interest to monitor thunderstorms and lightning due to the induced disruptions of its detectors.
	Furthermore, the Observatory offers opportunities for studying high-energy atmospheric phenomena such as Terrestrial Gamma-ray Flashes (TGFs).

	The Broadband Observatory of Lightning and TGFs (BOLT) is a lightning mapping array under construction,
	repurposing eleven former AERA stations
		from short air-shower radio measurements to multi-second VHF observations of lightning.
	The required FPGA and station DAQ modifications have been implemented, including freezing the sampling buffer and streaming large data volumes to external storage.

	Hardware for the array has been deployed between autumn 2024 and spring 2026.
	The three clusters are in different stages of commissioning, with the Core and Medium-range clusters recording first background data.

	Once fully commissioned,
		BOLT will provide the instrumentation to reconstruct the development of lightning above the Pierre Auger Observatory
		and correlate it with high-energy atmospheric phenomena, in particular TGFs.

\printbibliography

\end{document}